\documentclass[letterpaper,twocolumn,10pt]{article}
\usepackage{usenix2019_v3}
\newif\ifanon\anonfalse

\usepackage{tikz}
\usepackage{amsmath}
\usepackage{courier}
\usepackage{booktabs}
\usepackage{graphics}
\usepackage{graphicx}
\usepackage{layouts}
\usepackage{pgfplots}
\usepackage{pgf}
\usepackage{tabularx}

\usepackage[backend=bibtex, style=numeric, sorting=nty, maxnames=18]{biblatex}
\newcommand{\secref}[1]{\S \ref{#1}}

\begin{document}

\date{}


\title{\Large \bf You Really Shouldn't Roll Your Own Crypto: \\ An Empirical Study of Vulnerabilities in Cryptographic Libraries}

\author{
\ifanon
{\rm Anonymous}
\else
{\rm Jenny Blessing}\\
MIT
\and
{\rm Michael A. Specter}\\
MIT
\and
{\rm Daniel J. Weitzner}\\
MIT
\fi
} 

\maketitle

\begin{abstract}
The security of the Internet rests on a small number of open-source cryptographic libraries: a vulnerability in any one of them threatens to compromise a significant percentage of web traffic. Despite this potential for security impact, the characteristics and causes of vulnerabilities in cryptographic software are not well understood. In this work, we conduct the first comprehensive analysis of cryptographic libraries and the vulnerabilities affecting them. We collect data from the National Vulnerability Database, individual project repositories and mailing lists, and other relevant sources for eight widely used cryptographic libraries.

Among our most interesting findings is that only 27.2\% of vulnerabilities in cryptographic libraries are cryptographic issues while 37.2\% of vulnerabilities are memory safety issues, indicating that systems-level bugs are a greater security concern than the actual cryptographic procedures. In our investigation of the causes of these vulnerabilities, we find evidence of a strong correlation between the complexity of these libraries and their (in)security, empirically demonstrating the potential risks of bloated cryptographic codebases. We further compare our findings with non-cryptographic systems, observing that these systems are, indeed, more complex than similar counterparts, and that this excess complexity appears to produce significantly more vulnerabilities in cryptographic libraries than in non-cryptographic software.

\end{abstract}

\section{Introduction}
Cryptographic libraries are responsible for securing virtually all network communication, yet have produced notoriously severe vulnerabilities. In 2014, OpenSSL's Heartbleed vulnerability \cite{heartbleed} enabled attackers to read the contents of servers' private memory. More recently, in June 2020 GnuTLS suffered a significant vulnerability allowing anyone to passively decrypt traffic \cite{cve2020-13777}. Given the critical role these libraries play, a single vulnerability can have tremendous security impact. At the time of Heartbleed's disclosure, up to 66\% of all websites were vulnerable \cite{heartbleed}.

A common aphorism in applied cryptography is that cryptographic code is inherently difficult to secure due to its complexity; that one should not ``roll your own crypto.'' In particular, the maxim that complexity is the enemy of security is a common refrain within the security community. Since the phrase was first popularized in 1999 \cite{schneier-complexity}, it has been invoked in general discussions about software security \cite{chang2013your} and cited repeatedly as part of the encryption debate \cite{keysunderdoormats}. Conventional wisdom holds that the greater the number of features in a system, the greater the risk that these features and their interactions with other components contain vulnerabilities. 

Unfortunately, the security community lacks empirical evidence supporting the ``complexity is the enemy of security'' argument with respect to cryptographic software. Since software that implements cryptographic algorithms and protocols may have characteristics and testing processes that are quite distinct from non-cryptographic software, further study of the impact of complexity on security in cryptographic software specifically is needed.

In this paper, we conduct the first comprehensive, longitudinal analysis of cryptographic libraries and vulnerabilities affecting them. We examine eight of the most widely used cryptographic libraries and build a dataset of the 300+ entries in the National Vulnerability Database (NVD)~\cite{NVD} for these systems. In our analysis, we combine data from the NVD with information scraped from the projects' GitHub repositories, internal mailing lists, project bug trackers, and various other external references. We extensively characterize the vulnerabilities originating in cryptographic software, measuring exploitable lifetime, error type, and severity to better understand their security impact on cryptographic software.

Having surveyed the characteristics of the vulnerabilities themselves, we further investigate the causes of these vulnerabilities within cryptographic software. We study the relationship between software complexity and vulnerability frequency in an analysis of 11 years of vulnerabilities in cryptographic libraries, recording codebase size and cyclomatic complexity \cite{mccabe1976complexity} of each library. We also repeat our methodology on two non-cryptographic systems as a point of comparison.

Our findings include the following: just 27.2\% of vulnerabilities in cryptographic software are cryptographic issues as defined by the NVD, while 37.2\% of errors are related to memory management or corruption, suggesting that developers should focus their efforts on systems-level implementation issues. The median exploitable lifetime of a vulnerability in a cryptographic library is 4.18 years, providing malicious actors a substantial window of exploitation. At least one vulnerability is introduced for every thousand lines of code added in the most widely used cryptographic library, OpenSSL; and the rate of vulnerability introduction is up to three times as high in cryptographic software as in non-cryptographic software.

We hope that empirical evidence of security risk in cryptographic software will be helpful for software engineers working with cryptographic implementations. Our findings support the conventional wisdom preaching the dangers of ``rolling your own crypto'' and provide empirical estimates for cryptographic software risk. Through better understanding the causes and properties of vulnerabilities in cryptographic software, the security community can more effectively direct their efforts in mitigating against them.

Renewed government interest in regulating and modifying cryptographic code to facilitate law enforcement surveillance adds relevance and urgency to research on sources of vulnerabilities in cryptographic software. A greater technical understanding of the causes of security failures in cryptographic implementations and quantitative metrics for the consequences of the complexity of such systems will contribute to a more informed debate on the trade-offs between system security and national security when considering technical proposals to modify cryptographic protocols to provide law enforcement access to encrypted data, referred to as an ``exceptional access mechanism.''

The rest of the paper is organized as follows: We begin in \secref{section:related} with a brief discussion of related work examining cryptographic software, vulnerability life cycles, and the relationship between complexity and security. In \secref{section:methodology}, we describe our process for selecting systems and data sources in more detail and discuss mitigation strategies for inconsistencies in vulnerability data reporting. We then describe the properties of vulnerabilities in cryptographic libraries, measure the complexity of these libraries, and discuss its impact on security (\secref{section:analysis}). We conclude in \secref{section:discussion} and \secref{section:conclusion} with a discussion of takeaways for developers and users of cryptographic libraries and implications for the ongoing exceptional access debate.

\section{Related Work}
\label{section:related}

\paragraph{Relationship Between Complexity and Security:}
There have been several studies broadly investigating software complexity and its effect on security in general-purpose, non-cryptographic software. Ozment et al. \cite{ozment2006milk} conducted an empirical study of security trends in the OpenBSD operating system across approximately 8 years, focusing specifically on how vulnerability lifetimes and reporting rates have changed during that period.  They found that vulnerabilities live in the OpenBSD codebase for over two years on average before being discovered and patched. Zimmermann et al. \cite{zimmermann2010searching} similarly analyzed vulnerability correlation with various software metrics in the Windows Vista operating system, finding a weak correlation. More recently, Azad et al. \cite{azad2019less} studied the effects of software debloating in web applications, observing how removing significant percentages of the codebase in PHP web applications impacted vulnerabilities present. Their dataset included three cryptography-related vulnerabilities that were not removed through debloating, but their small sample size precludes drawing meaningful conclusions. In contrast to prior work, we focus specifically on cryptographic software and the vulnerabilities it produces.

\paragraph{Vulnerability Life Cycle:}
\label{section:related_lifecycle}
Lazar et al. \cite{lazar2014does} studied 269 cryptographic vulnerabilities, finding that only 17\% of the vulnerabilities they studied originated in the source code of cryptographic libraries, with the majority coming from improper uses of the libraries. In contrast to Lazar et al., we draw an important distinction between cryptographic software and cryptographic vulnerabilities: cryptographic vulnerabilities must necessarily originate in cryptographic software \footnote{The exception to this statement is when the vulnerability derives from an \textit{absence} of cryptography (i.e. data that should have been encrypted was not), but this is true as a general rule.}, but cryptographic software produces a broader class of vulnerabilities that are not only cryptographic in nature. Heartbleed, for instance, was a systems-level vulnerability caused by a missing bounds check. Lazar et al. studied only cryptographic vulnerabilities and therefore excluded Heartbleed and a significant percentage of vulnerabilities present in cryptographic software. 

Most recently, Walden \cite{walden2020impact} conducted a case study of OpenSSL in the aftermath of Heartbleed, analyzing changes made to the codebase after Heartbleed and observing that OpenSSL adopted additional recommended best practices, including reducing the complexity of the code, in the wake of the breach. While Walden also studied OpenSSL, they looked at more general software metrics than the granular complexity metrics studied in this paper. Walden counted vulnerabilities affecting each OpenSSL version instead of vulnerabilities introduced in each version, and only tracked the total size of the codebase rather than the individual sizes of the versions. Moreover, in this work we conduct a larger-scale analysis of codebase and vulnerability trends across multiple cryptographic libraries.

While there is a large existing body of work studying vulnerability life cycles and patches, prior work has not included cryptographic software in their datasets, despite the security-critical nature of such software. Li et al. \cite{li2017large} and Shahzad et al. \cite{shahzad2012large} both conducted large-scale analyses of vulnerability characteristics in non-cryptographic open-source software, including various operating systems and web browsers. Rescorla et al. \cite{rescorla2005finding} manually analyzed a dataset of 1,675 vulnerabilities from the Linux kernel and other large software systems, defining vulnerability lifetime through window of exposure, and found inconclusive evidence that public vulnerability disclosure is worth the security implications. The unique characteristics of cryptographic software suggest that vulnerability data from non-cryptographic systems may not be applicable, necessitating a separate investigation of cryptographic software specifically.

\section{Methodology}
\label{section:methodology}
To study the causes and characteristics of vulnerabilities in cryptographic software, we collect source code repository and vulnerability data from eight open-source cryptographic libraries. We chose to look at open-source cryptographic libraries because these systems are the most widely used software codebases where the source code is almost entirely cryptographic in nature (i.e. implements cryptographic primitives, protocols, or algorithms). We define vulnerability as an entry in the Common Vulnerabilities and Exposures (CVE) list maintained by MITRE \cite{CVE}.

For each cryptographic system, we measured (1) source code repository size and cyclomatic complexity over time and (2) vulnerabilities published in the NVD over the same time period. For each vulnerability, we measured (1) error type as reported in the NVD, (2) severity, and (3) exploitable lifetime. To investigate whether trends observed are unique to cryptographic source code, we collect the same complexity metrics from three systems that are non-cryptographic in nature.

While we collected some of the repository and vulnerability data through automated web scraping, due to inconsistencies and inaccuracies in the NVD and other sources much of our dataset was manually compiled and analyzed. We describe issues encountered with the quality of the NVD in more detail in \secref{section:data_sources}. \secref{section:systems} and \secref{section:complexity_metrics} describe our criteria for selecting systems and complexity metrics, and ~\secref{section:lifetime} discusses the challenges of determining system versions affected by a vulnerability when calculating vulnerability lifetime.

\subsection{Vulnerability Data Sources}
\label{section:data_sources}

We use the National Vulnerability Database (NVD) \cite{NVD}, managed by the National Institute of Standards and Technology (NIST), as a standardized source of vulnerabilities from which to construct our dataset. When a new CVE (Common Vulnerabilities and Exposures) ID \cite{CVE} is created, the NVD calculates a severity score (CVSS) \cite{CVSS} and performs additional analysis before adding the vulnerability to the NVD. While individual product bug trackers often provide more granularity, they do not allow us to standardize or compare across systems as the NVD does.

We scrape CVE data from two third-party platforms, CVE Details \cite{cvedetails} and OpenCVE \cite{opencve}, which contain much the same data as the official NVD but organize CVEs by product and vendor, enabling us to retrieve all CVEs for a particular system. OpenCVE is a newer replacement for CVE Details, which at the time of data collection had not been updated since 2019. We began our study prior to the release of OpenCVE in December 2020, and so we include data scraped from both CVE Details and OpenCVE.

To supplement the data provided by the NVD, we manually retrieve CVE information from individual projects' bug trackers, mailing lists, blogs, and other external references. Incorporating information from project-specific sources and standardized external sources enables us to conduct an analysis that is sufficiently in-depth to generalize across multiple projects.

In total, our dataset consists of $n = 312$ CVEs in cryptographic libraries and 2,000+ CVEs in non-cryptographic software. In our study of cryptographic vulnerability characteristics, we consider only CVEs published by the NVD between 2010 and 2020, inclusive.

\subsection{Systems Analyzed}
\label{section:systems}
We select cryptographic and non-cryptographic systems for inclusion in our study on the basis of the following requirements:

\begin{enumerate}
    \item \textbf{Open-Source:} A critical component of this work is measuring software characteristics of codebases at particular points in time, and so we consider only systems where we have access to the source code.
    \item \textbf{Written in C/C++:} To ensure an accurate comparison, we select only systems primarily written in C or C++. For instance, we exclude Bouncy Castle from our list of cryptographic libraries because its primary distribution is written in Java \cite{bouncycastle-java}. Prior work \cite{gaynor2020memory} has demonstrated significant differences in vulnerability causes in memory-unsafe C/C++ source code compared to systems written in memory-safe languages such as Java.
    \item \textbf{Sufficient CVE Reporting:} We further select for systems that have sufficient quantities of CVEs reported to allow us to generalize. In cryptographic software, we consider only cryptographic libraries that have at least 10 CVEs published from 2010 - 2020. In non-cryptographic software, to ensure that we avoid systems which are under-reporting CVEs we select only non-cryptographic systems with comparatively high absolute vulnerability counts relative to all products in the NVD. Both non-cryptographic systems we collect CVE data from are among the top 10 software products with the largest number of CVEs as studied by Li et al. \cite{li2017large}.
\end{enumerate}

In line with the criteria above, we study the following six cryptographic libraries: OpenSSL \cite{openssl}, GnuTLS \cite{gnutls}, Mozilla Network Security Services (NSS) \cite{nss}, Botan \cite{botan}, Libgcrypt \cite{libgcrypt}, and WolfSSL \cite{wolfssl}. We additionally collect data from the LibreSSL \cite{libressl} and BoringSSL \cite{boringssl} libraries, though they do not meet our CVE reporting requirement, as part of a case study of OpenSSL forks discussed further in \secref{section:open_boring_libre}.

We also study three non-cryptographic systems, Ubuntu Linux \cite{ubuntu}, Chrome \cite{chrome}, and Wireshark \cite{wireshark}, as part of our comparison of the security impact of complexity. We are limited in the number of non-cryptographic systems we can consider by the quality of the CVE data available. In order to accurately track vulnerabilities within a system, we need software systems that report all product versions affected for each CVE, not merely the patched version. This level of granularity gives us the version in which a CVE was introduced, which we use to calculate both exploitable lifetime and individual version security. As we discuss further in \secref{section:cve_reporting}, few systems consistently report CVE version data.

\begin{table}[]
\centering
\begin{tabular}{@{}cc|c@{}}
\toprule
\multicolumn{1}{l}{}    & \textbf{Cryptographic Library} & \textbf{Num. of CVEs} \\ \midrule
\multicolumn{1}{c|}{1.} & OpenSSL                        & 153                   \\
\multicolumn{1}{c|}{2.} & GnuTLS                         & 43                    \\
\multicolumn{1}{c|}{3.} & Mozilla NSS                    & 40                    \\
\multicolumn{1}{c|}{4.} & WolfSSL                        & 36                    \\
\multicolumn{1}{c|}{5.} & Botan                          & 21                    \\
\multicolumn{1}{c|}{6.} & Libgcrypt                      & 12                    \\
\multicolumn{1}{c|}{7.} & LibreSSL                       & 6                    \\
\multicolumn{1}{c|}{8.} & BoringSSL                      & 1                    \\ \midrule
\multicolumn{1}{l}{}    & \textbf{Total}                 & \textbf{312}          \\ \bottomrule
\end{tabular}
\caption{The eight cryptographic libraries studied, listed in order of the total CVEs published in each library from 2010 through 2020.}
\label{tab:cves}
\end{table}

\subsection{Complexity Metrics Analyzed}
\label{section:complexity_metrics}
There are a variety of mechanisms for approximating software complexity. We select two particular complexity metrics through which to study security outcomes across different systems: total lines of code (LOC) and cyclomatic complexity. Prior work \cite{zimmermann2010searching, shin2010evaluating} has shown that these two metrics are among the best complexity predictors of vulnerabilities in non-cryptographic software. We define cyclomatic complexity as the number of linearly independent paths through a system's source code, following McCabe's 1976 definition~\cite{mccabe1976complexity}.

\medskip
\noindent\textbf{Lines of Code:} We use \texttt{cloc}~\cite{danial2009cloc}, a command-line tool, to count the total lines of code for each language in a codebase. Throughout our study, we only consider C or C++ source code lines in our count, and we exclude blank lines, comment lines, and header files. We collect LOC measurements of the relevant systems over time as well as for specific version releases.

\medskip
\noindent \textbf{Cyclomatic Complexity:} We use a separate command-line tool, \texttt{lizard} \cite{yin2020lizard}, to calculate cyclomatic complexity of all C and C++ source files. Lizard calculates the complexity of each file individually and averages them together, outputting a single average cyclomatic complexity number (CCN).

\subsection{Calculating Vulnerability Lifetime}
\label{section:lifetime}

Determining how long a vulnerability was actively exploitable, rather than merely how long it existed in the source code, is critical to fully understanding its security impact. The exploitable lifetime represents the period in which the vulnerability could actually be abused, while the codebase lifetime represents how long it took for someone to notice the vulnerability. We therefore define a vulnerability's lifetime as the period of time in which it can be exploited by a malicious actor. A vulnerability is exploitable from the moment the source code in which it lives is released until the patch for the vulnerability is installed by an affected client. Calculating this lifetime, then, involves determining the release date of the first version affected and the release date of the patch. Because clients frequently continue using outdated versions without updating, even in the wake of a major security breach \cite{durumeric2014matter}, these calculations represent a lower bound on the actual exploitable lifetime.

\subsubsection{Determining Versions Affected}
\label{subsection:versions}

The first step in calculating a vulnerability's exploitable lifetime is to determine which versions of a system the vulnerability affected. The NVD's reporting requirements ask that CVE listings include either the patch version or the full list of affected versions. Because the patch version is usually the latest version of the system, it is much easier to obtain and report, and so the vast majority of CVEs contain only the patch version.

We manually review systems and the CVEs reported in them to ensure we only collect data from systems that consistently and accurately record affected versions. For each CVE ID, CVE Details provides a list of versions affected in a tabular format. In our review, we encounter several common types of errors with reporting of affected versions:
\begin{enumerate}
    \item \textbf{No Versions Listed:} A CVE had only a special character or blank space listed as the affected version, indicating that no versions related to the CVE are known, as in CVE-2018-16868 in GnuTLS \cite{cve2018-16868}. We necessarily exclude these vulnerabilities from lifetime calculations.
    \item \textbf{One Version Listed:} A significant percentage of CVEs had only one affected version listed. After manually reviewing several hundred CVEs and patches, we conclude that this is almost always in error and represents a situation where only the current version (as of CVE discovery) is listed instead of all affected versions. We eliminate such vulnerabilities from our dataset, since to include them would artificially lower the lifetime data collected.
    \item \textbf{All Versions Listed:} In some systems, the text description of the CVEs gives the affected versions as some variation of ``version X and before'', and the table containing affected versions lists all versions the system has ever released. While the table produced is a technically correct reading of the CVE description, we found that this reporting is almost always incorrect and should not be interpreted literally. Our approach here differs from Rescorla \cite{rescorla2005finding}, who also measured exploitable lifetime and collected affected version data but interpreted phrasing like ``version X and earlier'' at face value. In our work, we avoid studying any systems that consistently exhibit this version reporting pattern.
\end{enumerate}

After thorough manual review, we identified two cryptographic systems (OpenSSL and GnuTLS) and two non-cryptographic systems (Ubuntu Linux and Wireshark) that have sufficiently accurate version reporting to calculate lifetime. For each of these systems, we scrape all affected versions from CVE Details and sort the versions alphanumerically to obtain the first affected version and the patch version. In the case of OpenSSL, the project itself keeps a detailed record of vulnerabilities and affected versions, so we scraped from the project's vulnerabilities page \cite{openssl_vulnerabilities} instead. While Mozilla NSS does not report affected versions for its CVEs, we use the CVE references to manually estimate versions affected.

\subsubsection{Mapping Versions to Release Dates}
Once we know the initial and patch versions of a CVE, we need to determine the release dates of those versions in order to calculate lifetime. For each system we study, we construct a dataset of versions and release dates by manually reviewing individual system websites and developer mailing lists for version release dates. While most systems clearly publish the release dates of major versions, we found that minor version release dates were trickier to track down, particularly for versions released over a decade ago, and required substantial manual trawling of various mailing lists. In a small number ($n = 3$) of cases, we were only able to find the month and year of a vulnerability's release date. In these cases, we used the 15th of the month as an approximation of the release date.

Given the time-consuming nature of determining release dates for all versions of a system and the difficulty of finding systems that accurately report affected versions for each vulnerability, we limited ourselves to collecting CVE lifetime data from three cryptographic libraries (OpenSSL, GnuTLS, and Mozilla NSS) and two non-cryptographic systems (Ubuntu and Wireshark).

\subsection{CVE Reporting Practices}
\label{section:cve_reporting}
Since our analysis relies heavily on the quality of the vulnerability reporting in the National Vulnerability Database, we briefly discuss observed reporting practices in both cryptographic and non-cryptographic systems.

\medskip
\noindent \textbf{How reliable is the CVE data on cryptographic libraries?}
After manually reviewing CVEs reported in approximately ten of the most widely used cryptographic libraries, we observe that OpenSSL has a far greater number of CVEs than any other cryptographic library, with 153 CVEs published during our timeframe of 2010 - 2020 compared to the second-highest count of 43 CVEs in GnuTLS. The question, then, is whether this difference is due to variations across libraries in security, attention, internal reporting policies, or some linear combination of the three.

We examine the CVE data on the LibreSSL and BoringSSL forks of OpenSSL as a test case to study how the official NVD vulnerability counts compare. While OpenSSL's absolute vulnerability count from 2010 - 2020 was 153 CVEs, LibreSSL and BoringSSL recorded just six and one CVEs, respectively, across the same time period. Since we conducted extensive manual examination of project commits, team mailing lists, and security advisories as part of our case study of the forks (described in greater detail in \secref{section:open_boring_libre}), we confirm that the NVD count of vulnerabilities affecting LibreSSL and BoringSSL is substantially lower than the actual number of vulnerabilities affecting those libraries. We discuss this discrepancy further in \secref{subsubsection:cves_removed}. This finding reinforces the unreliability of absolute vulnerability counts as an indication of project security.

\medskip
\noindent \textbf{How do CVE reporting practices vary between cryptographic and non-cryptographic systems?}
Overall, CVE reporting quality is more consistent and accurate among cryptographic software and systems focused on security (such as Wireshark and Tor). Of the top 10 software products with the highest number of reported CVEs as calculated by Li et al. \cite{li2017large}, after manual review we find that only 3 of them---Ubuntu, Wireshark, and Android---consistently report affected system versions. Most projects report only the version in which the CVE was patched.

Non-cryptographic projects more frequently report a skewed distribution of vulnerability severity as determined by the standard CVE severity scoring system, CVSS. For instance, we find that Android’s average and median CVSS score for CVEs published from 2012 through 2020 were 6.88 and 7.2, respectively, suggesting internal practices that favor reporting only higher-severity vulnerabilities. By contrast, median severity in cryptographic libraries is consistently 5.0, as discussed in \secref{section:cve_severity}.

In summary, we observe the following common issues with CVE reporting across cryptographic and non-cryptographic systems:
\begin{enumerate}
    \item The number of CVEs reported in a project is small relative to the widespread use of the project. We expected this of non-cryptographic systems, but found this is also true of a surprising number of cryptographic libraries. For instance, only one CVE has ever been reported in cryptlib \cite{cryptlib}, a commonly used library.
    \item CVEs reported skew towards high-severity vulnerabilities, suggesting vulnerabilities reported as CVEs are being somewhat cherry-picked by the vendor.
    \item Affected versions are frequently not reported, making it difficult to find the version in which a CVE was introduced.
\end{enumerate}

\subsection{Limitations}

\subsubsection{National Vulnerability Database (NVD)}

\textbf{Reporting Bias:} The NVD suffers from selection bias in that not all systems report vulnerabilities when they are discovered. Some vendors pay little attention to the CVE database and do not bother to register vulnerabilities as CVEs, and others skew towards only reporting high-severity CVEs.

We mitigate this by intentionally selecting only non-cryptographic systems with comparatively high vulnerability counts. We further avoid using known CVE count as an absolute metric in our analysis because of these limitations, since this may be a better indicator of a vendor's policy than of a system's security.

\medskip
\noindent \textbf{Quality Bias:} Even when a system has high quantities of CVEs reported, the CVE listings often fail to include sufficient detail. For instance, as discussed in \secref{subsection:versions}, most systems do not accurately report versions affected by a CVE.

\subsubsection{Systems Studied}

\textbf{Open-Source:} All the systems we study are open-source projects. It is possible that the trends we observe will not be present in proprietary, closed-source software.  In order to accurately measure complexity, though, we find it necessary to focus solely on open-source systems.

\medskip
\noindent \textbf{Non-Technical Factors:} The security of a system is impacted by many economic and human factors in addition to codebase complexity and other software metrics.  Software development and testing practices, developer experience level, and other considerations all affect the quantity of vulnerabilities introduced but are not reflected in codebase data.

\section{Results}
\label{section:analysis}
Here, we present our empirical analysis of the causes and characteristics of vulnerabilities in cryptographic software. We begin by exploring qualitative and quantitative properties of vulnerabilities discovered in the cryptographic libraries we study. Next, we study how cryptographic software complexity affects how many of these CVEs are present in a codebase. Finally, we look at how the relationship between complexity and vulnerability count in cryptographic software compares with the same effect in non-cryptographic software, finding that cryptographic software complexity is indeed more dangerous than non-cryptographic software complexity. In the interest of open access, we intend to publicly release all data used in this analysis.

\begin{figure}
\begin{center}
    \includegraphics[scale=0.25]{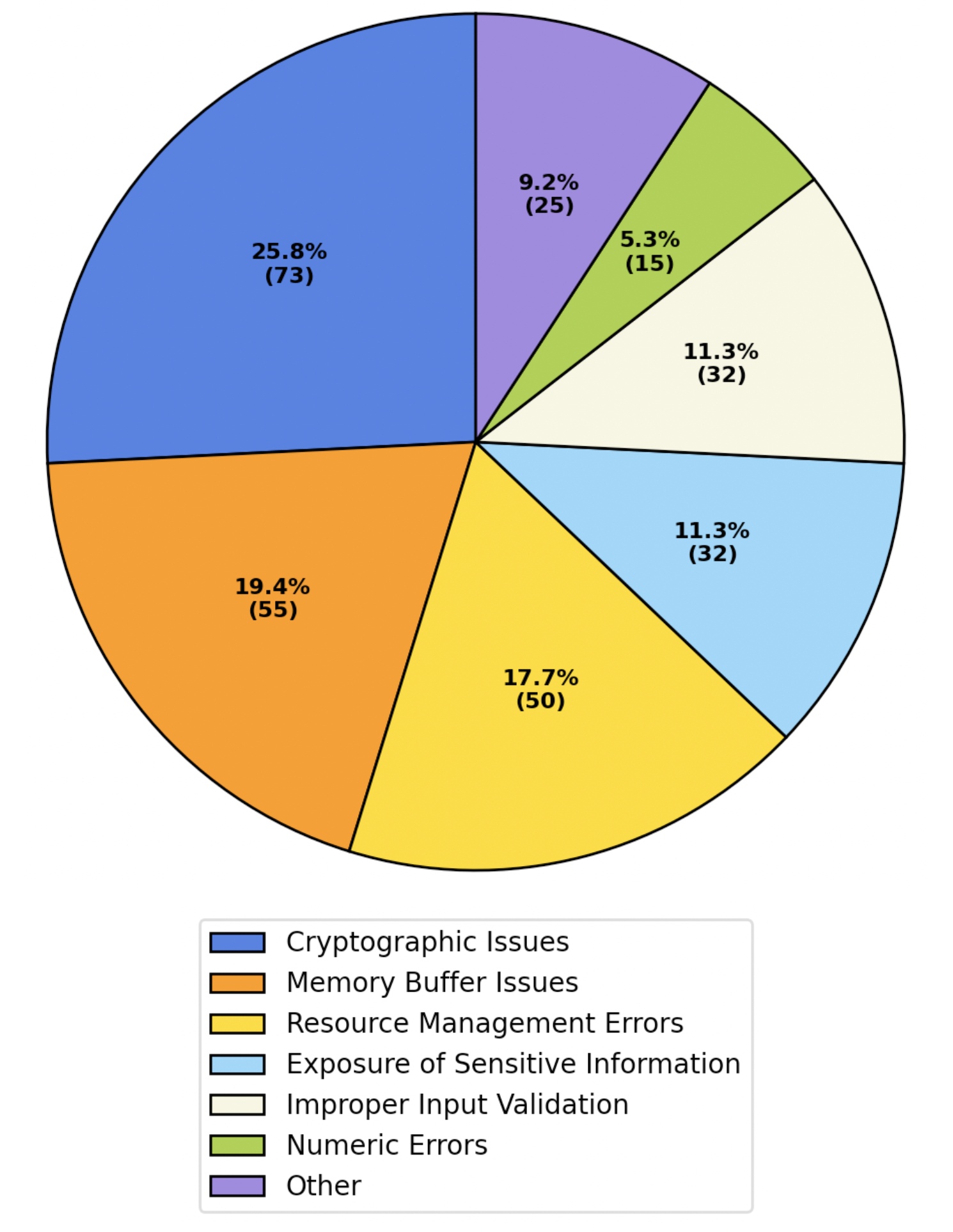}
\end{center}
\caption{Breakdown of vulnerability types according to CWE description for all cryptographic software vulnerabilities studied.}
\label{fig:cwe_plot}
\end{figure}

\subsection{Characteristics of Vulnerabilities in \\Cryptographic Software}
\label{section:cve_characteristics}
In this section, we characterize the 312 vulnerabilities affecting the eight cryptographic libraries studied. We first categorize vulnerabilities by type, including broadly categorizing them as cryptographic or non-cryptographic as defined in \secref{subsubsection:crypto_cves}, and also study the severity and exploitable lifetime of these vulnerabilities.

\subsubsection{Vulnerability Type}
Vulnerabilities in cryptographic software can be divided into several different categories based on error type. The National Vulnerability Database assigns each vulnerability a Common Weakness Enumeration (CWE) \cite{CWE} broadly classifying the vulnerability type. Of the 312 CVEs in our dataset, 38 do not have a CWE listed, so we exclude them in this section for a new total of $n = 274$ CVEs.

Because CWE labeling can be inconsistent and overly granular, particularly when studied across multiple systems and years, we manually group CWEs into broader categories. For instance, CWE-125: Out-of-bounds Read, CWE-787: Out-of-bounds Write, and CWE-131: Incorrect Calculation of Buffer Size can all be grouped under the larger category of Memory Buffer Issues. The 312 vulnerabilities we studied had 36 distinct CWEs, so we combined these 36 CWEs into 7 more general categorizations.

Figure \ref{fig:cwe_plot} shows each of the 7 categories created, the absolute count of CWEs in each category, and their relative percentages \footnote{Note that a small number of CVEs have more than one CWE listed, and so the 282 total CWEs shown in the figure is slightly higher than the 274 CVEs with CWEs included in this experiment.}. While cryptographic issues are the largest individual category, comprising 25.8\% of CWEs, memory-related errors are the most common overall type, producing 37.1\% of CWEs when combining memory buffer issues and resource management errors. A further 27.9\% of CWEs arise from various smaller sub-categories, including exposure of sensitive information, improper input validation, and numeric errors (i.e. errors in numerical calculation or conversion).

\subsubsection{\textbf{Cryptographic Vulnerabilities}}
\label{subsubsection:crypto_cves}

\textbf{How many vulnerabilities in cryptographic libraries are actually cryptographic?} We previously distinguished between cryptographic software and cryptographic vulnerabilities in \secref{section:related_lifecycle}. Cryptographic software can produce plenty of vulnerabilities where the error is non-cryptographic in nature, as shown in Figure~\ref{fig:cwe_plot}.

We define \emph{cryptographic vulnerability} more broadly than Lazar et al. \cite{lazar2014does} did in their study of cryptographic CVEs. They considered only CVEs tagged as ``CWE-310: Cryptographic Issue'' under the NVD categorization, which is the largest cryptographic CWE category. However, the NVD also contains a number of more specific CWE classifications, such as CWE-326: Inadequate Encryption Strength or CWE-327: Use of a Broken or Risky Cryptographic Algorithm, that should also broadly be considered cryptographic issues. We find six such CWE categories \footnote{We consider the following CWEs to be cryptographic: CWE-310, CWE-327, CWE-326, CWE-320, CWE-335, CWE-330, CWE-311.} (representing 25 CVEs) and include CVEs classified under them in our accounting of cryptographic issues. For CVEs with multiple CWEs, if at least one of the CWEs is cryptographic then we consider the CVE to also be cryptographic.

Our findings show that just 27.24\%, or approximately 1 in 4, CVEs in cryptographic software are actually cryptographic. Table \ref{tab:cryptocves} shows the breakdown of cryptographic CVEs in each of the six cryptographic libraries we study. The percentages are remarkably consistent across libraries, with the exception of Botan where only 10.52\% of CVEs are cryptographic.

\begin{table*}[t]
\centering
\begin{tabular}{@{}cc|c|c|cl@{}}
\toprule
\multicolumn{1}{l}{} & \textbf{System} & \textbf{Num. of CVEs with CWEs} & \textbf{Num. of Cryptographic CVEs} & \textbf{\% Cryptographic} &  \\ \midrule
\multicolumn{1}{c|}{1.} & OpenSSL     & 134 & 38 & 28.35\% &  \\
\multicolumn{1}{c|}{2.} & GnuTLS      & 39  & 12 & 30.76\% &  \\
\multicolumn{1}{c|}{3.} & Botan       & 19  & 2  & 10.52\%  &  \\
\multicolumn{1}{c|}{4.} & Libgcrypt   & 12  & 3  & 25\%    &  \\
\multicolumn{1}{c|}{5.} & WolfSSL     & 32  & 9  & 28.13\% &  \\
\multicolumn{1}{c|}{6.} & Mozilla NSS & 32  & 9 & 28.13\% &  \\ \midrule
\multicolumn{1}{l}{} & \textbf{Total}  & \textbf{268}                    & \textbf{73}                         & \textbf{27.24\%}          &  \\ \bottomrule
\end{tabular}
\caption{Percentage breakdown of cryptographic versus non-cryptographic CVEs in cryptographic libraries. We exclude LibreSSL and BoringSSL from this table due to their small vulnerability counts.}
\label{tab:cryptocves}
\end{table*}

\begin{table*}[t]
\centering
\begin{tabular}{@{}cc|c|c|c|c@{}}
\toprule
\multicolumn{1}{l}{} & \textbf{System} & \textbf{Num. CVEs} & \textbf{Median Lifetime} & \textbf{Avg. Lifetime} & \textbf{StdDev Lifetime} \\ \midrule
\multicolumn{1}{c|}{1.} & OpenSSL     & 143 & 4.14 & 6.16 & 3.19 \\
\multicolumn{1}{c|}{2.} & GnuTLS      & 22  & 1.65  & 2.35  & 2.18  \\
\multicolumn{1}{c|}{3.} & Mozilla NSS & 36  & 12.43 & 9.43    & 5.85 \\ \midrule
\multicolumn{1}{l}{} & \textbf{Total}  & \textbf{201}       & \textbf{4.18}          & \textbf{5.13}       & \textbf{4.29}         \\ \bottomrule
\end{tabular}
\caption{Exploitable lifetimes (in years) of vulnerabilities in cryptographic libraries. We only include OpenSSL, GnuTLS, and Mozilla NSS in our calculations since only these three libraries report data on versions affected by vulnerabilities.}
\label{tab:lifetimes}
\end{table*}

\begin{table*}[]
\centering
\begin{tabularx}{.9\textwidth}{>{\raggedright\arraybackslash}c>{\centering\arraybackslash}c>{\centering\arraybackslash}c>{\centering\arraybackslash}c>{\centering\arraybackslash}c>{\centering\arraybackslash}cc}
\toprule
\textbf{Major Version} & \textbf{Release Date} & \textbf{\begin{tabular}[c]{@{}c@{}}Average\\ CCN\end{tabular}} & \textbf{\begin{tabular}[c]{@{}c@{}}Most Recent\\ Minor Version\end{tabular}} & \textbf{LOC Change} & \textbf{\begin{tabular}[c]{@{}c@{}}CVEs\\ Introduced\end{tabular}} & \textbf{\begin{tabular}[c]{@{}c@{}}CVEs /\\ KLOC\end{tabular}} \\
\midrule
1.0.2 & 1/22/2015 & 6.9 & 1.0.1l & 22,236 & 25 & 1.12 \\
1.0.1 & 3/14/2012 & 6.7 & 1.0.0h & 18,766 & 33 & 1.76 \\
1.0.0 & 3/29/2010 & 6.7 & 0.9.8n & 15,510 & 12 & .77  \\
0.9.8 & 7/5/2005 & 6.8 & 0.9.7g & 23,174 & 28 & 1.2   \\
\bottomrule
\end{tabularx}
\caption{CVEs introduced per thousand lines of C/C++ source code (KLOC) in four versions of OpenSSL.}
\label{tab:openssl_versions}
\end{table*}

\medskip
\noindent \textbf{Of high-severity vulnerabilities, how many are cryptographic?} We also investigate the prevalence of cryptographic CVEs in high-severity vulnerabilities specifically. CVSS v2.0 defines vulnerabilities with a score of 7.0 - 10.0 to be ``high'' in severity \cite{cvssv2v3}, so we exclude CVEs not in that range. We find that 64 of the 274 CVEs with CWEs are classified as high-severity under CVSS v2.0. Of these 56 CVEs, 2 had a cryptographic CWE and 54 were non-cryptographic. Of the most severe CVEs, just 3.57\% were cryptographic, a substantially lower percentage compared to 27.24\% of all CVEs.

Since we observed a significant difference in cryptographic vulnerability frequency in severe vulnerabilities compared to the total vulnerability population under CVSS 2.0 scoring, we also examine high-severity vulnerabilities under CVSS 3.0, a more recent CVSS scoring system released in 2015. CVSS v3.0 further divides severe vulnerabilities into two sub-categories: ``high'' for scores in the range of 7.0 - 8.9, and ``critical'' for scores between 9.0 - 10.0. Due to qualitative changes in the scoring system from v2.0 to v3.0, there is a substantially larger number of severe vulnerabilities: 99 CVEs are either ``high'' or ``critical.'' Of these 99 CVEs, 11 had a cryptographic CWE and 88 had only non-cryptographic CWEs. We calculate that 11.11\% of severe vulnerabilities are cryptographic.

Overall, while 27.24\% of all vulnerabilities in cryptographic libraries are cryptographic errors, only 3.57\% to 11.11\% of the most severe vulnerabilities are errors in the cryptographic nature of the code.

\subsubsection{Vulnerability Lifetime}
\label{subsubsection:crypto_lifetimes}

We calculate a vulnerability's ``exploitable'' lifetime by measuring time elapsed between the version in which the vulnerability was introduced and the version where the vulnerability was patched, as discussed in \secref{section:lifetime}. Of the eight cryptographic libraries considered in this study, we find that only three of them (OpenSSL, GnuTLS, and Mozilla NSS) report data on versions affected for at least one-third of their CVEs, and so we calculate lifetimes for these systems only. Within these three libraries, we further remove CVEs that did not report affected versions or did not do so accurately, giving us a total sample size of 201 vulnerabilities.

Table \ref{tab:lifetimes} displays the median and average lifetimes for each system along with the sample standard deviation. We find that overall, the median lifetime of a vulnerability in cryptographic software is 1,527 days, or 4.18 years. In all three systems, the average lifetime is slightly greater than the median due to a small number of outlier CVEs that persist in the software for abnormally long periods. Because these calculations necessarily include only vulnerabilities that have been discovered and reported, these numbers should be interpreted as lower bounds on the actual lifetimes.

\subsubsection{Vulnerability Severity}
\label{section:cve_severity}
We further study vulnerability severity across systems using CVSS v2 scores. While the latest CVSS v3 is the most recent CVSS version, the NVD's policy was not to retroactively score vulnerabilities published prior to December 20, 2015 according to the CVSS v3.0 scale \cite{cvss_scoring}, so a number of vulnerabilities in our dataset only have v2 scoring. To maintain consistency, we use CVSS v2 for all CVEs.

Across the eight cryptographic libraries studied and the 312 CVEs affecting them, the average severity score was 5.21, with a standard deviation of 1.73. All libraries except Libgcrypt had a median severity of 5.0, with Libgcrypt's median severity slightly lower at 4.3. There was very little variation among libraries in vulnerability severity. We conclude that an average severity score of around 5 is a sign of healthy CVE reporting and scoring. For a severity score to be higher would suggest that the system is under-reporting CVEs by reporting only the more severe vulnerabilities.

\subsection{Code Complexity and Security}
\label{section:complexity}
Excessive complexity is often cited within the security community as the reason for adverse security outcomes. We explore this association by studying how the size of a codebase affects vulnerabilities introduced in two different studies. First, we study how many CVEs are introduced for every thousand lines of code across relevant OpenSSL versions. Second, we use the LibreSSL and BoringSSL forks of OpenSSL as a natural experiment to study how the changes made in the forks impacted the security of the projects.

\subsubsection{Correlation Between LOC and Vulnerability Count}
To control for variations in vulnerability reporting practices, rather than considering entirely different systems we instead contrast CVEs introduced across different versions of the same system, OpenSSL. We consider only OpenSSL here because it is the only cryptographic library with sufficient quantity and quality of CVE entries to allow us to do such a comparison. It is also to date still the single most widely used cryptographic library \cite{censys}.

Here, we examine whether there is a linear correlation between the lines of code introduced in a version and the number of CVEs introduced. We select four OpenSSL versions (0.9.8, 1.0.0, 1.0.1, and 1.0.2) whose release dates roughly span the 10-year period from 2005 to 2015. OpenSSL 0.9.8 was released in July 2005, and 1.0.2 was released in January 2015. Versions released within this timespan are old enough that vulnerabilities have had time to be discovered but recent enough that the results are still relevant for contemporary software development. Since OpenSSL releases major versions every two to three years on average, we are necessarily limited in the number of versions we can include in our study.

For each of the four releases, we approximate the net lines of code added in the version by measuring the overall size of the major version in question and the most recent prior version (in all four cases, a minor version) and taking the difference. The source code for all versions was obtained from the releases stored in OpenSSL’s source code repository \cite{openssl_releases}. A lack of data on the precise commits that were included in a version release makes it necessary to measure LOC added in a version in this indirect manner, but the LOC difference between a release and the one immediately preceding it gives a very near approximation of the size of the version. It should be noted that this calculation yields the net change of lines added and removed, rather than solely lines added, which represents a better approximation of the impact of the version on the codebase. Again, we consider only C \& C++ source code and exclude blank lines, comment lines, and header files.

Column 5 in Table \ref{tab:openssl_versions} gives the LOC change for each version, for an average of 19,921.5 lines of C added per version. We further calculate the number of CVEs introduced in each version using vulnerability data tracked by the OpenSSL project \cite{openssl_vulnerabilities}. We estimate the number of CVEs introduced per thousand lines of code by taking the ratio of columns 5 and 6 in Table \ref{tab:openssl_versions}.

Column 7 shows that, on average, around 1 CVE is introduced in OpenSSL for every thousand lines of code added. This ratio should be interpreted as a lower bound since this necessarily includes only vulnerabilities that have been discovered. The ratio of vulnerabilities existing in the codebase per thousand LOC is very likely higher when taking into account vulnerabilities that have not yet been discovered, though it is impossible to know just how much higher. Furthermore, we made a methodological decision to calculate these sizes based on the net LOC difference (which takes into account LOC removed) instead of solely considering LOC added, since this gives a more complete accounting of changes made in the version.

\begin{figure}[h]
\begin{center}
    \scalebox{0.5}{\input{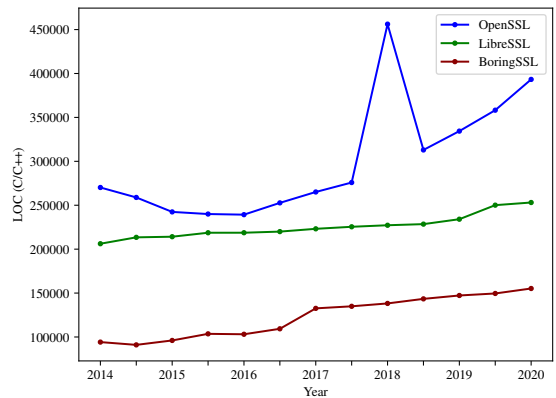}}
\caption{Relative sizes of OpenSSL, LibreSSL, and BoringSSL over seven years from July 2014 through July 2020. LOC count includes only C and C++ source code (excluding blank lines, comment lines, and header files). The unusual 2018 spike in the size of OpenSSL's codebase was due to testing files temporarily added for FIPS-140 compliance~\cite{openssl_spike}.}
\label{fig:openboringlibre_size}
\end{center}
\end{figure}

Comparing across versions, we observe a consistent trend of larger version sizes producing higher quantities of CVEs. The outlier to the overall trend is version 1.0.1, which is the version in which Heartbleed was introduced. OpenSSL's codebase received a sudden influx of attention in 2014 after Heartbleed was discovered, which may partially explain the higher number of vulnerabilities discovered. Spearman's non-parametric correlation test calculates a correlation coefficient of $\rho = 0.8$, indicating a strong linear relationship.

\subsection{Case Study: OpenSSL, LibreSSL, and BoringSSL}
\label{section:open_boring_libre}
The Heartbleed vulnerability \cite{heartbleed} gained international attention in April 2014, bringing OpenSSL into the spotlight with it. The increased scrutiny of the OpenSSL codebase in the wake of Heartbleed prompted the creation of two major forks of the codebase: LibreSSL, developed by the OpenBSD project and released on July 11, 2014 \cite{libre_releases}, and BoringSSL, developed by Google and released on June 20, 2014 \cite{langley2015}.

Although LibreSSL and BoringSSL are both forks of OpenSSL, they were intended for very different purposes: LibreSSL was conceived of as a replacement for OpenSSL that maintained prior API compatibility and portability \cite{libressl}, while BoringSSL was developed for internal Google use only. OpenBSD forked OpenSSL with the goal of creating a more modern and secure TLS library after a particular disagreement over the way OpenSSL handled memory management \cite{libressl, libre_beck}. The BoringSSL project, on the other hand, specifically states that the library is not recommended for external use outside of Google \cite{boringssl}. This difference in purpose helps to explain why BoringSSL diverges more from OpenSSL than LibreSSL in overall size (as shown in Figure \ref{fig:openboringlibre_size}) and features offered.

\subsubsection{Code Removal}
\label{subsubsection:code_removal}

Post-fork, LibreSSL and BoringSSL both removed significant amounts of the OpenSSL codebase. On April 7, 2014, the day that Heartbleed was patched and announced, OpenSSL contained 269,179 lines of C and C++ source code. In the months that followed from early April through June 2014, LibreSSL removed roughly 60,000 C/C++ LOC while BoringSSL removed 180,000 LOC.

Figure \ref{fig:openboringlibre_size} shows a comparison of the codebase sizes over time, beginning in July 2014 once all three libraries had been released. The comparatively large size of OpenSSL compared to the other two is primarily due to OpenSSL's maintenance of legacy ciphers and protocols in order to maintain backwards compatibility and portability.

\begin{figure}
\begin{center}
    \scalebox{0.5}{\input{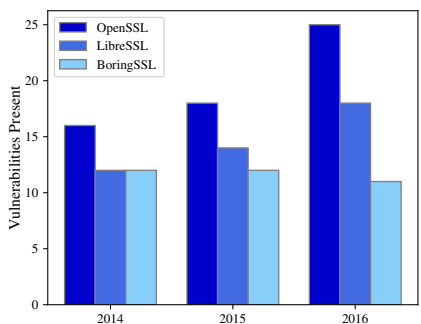}}
\caption{Vulnerabilities discovered in OpenSSL after the initial releases of LibreSSL and BoringSSL with a comparison of how many of those vulnerabilities also affected LibreSSL and BoringSSL.}
\label{fig:openboringlibre_cves}
\end{center}
\end{figure}

We summarize the major changes made by LibreSSL and BoringSSL in the immediate aftermath of Heartbleed below, focusing on features removed between April and July of 2014.

\paragraph{LibreSSL Changes} The OpenBSD team built LibreSSL under the design that the library would only be used on a POSIX-compliant OS with a standard C compiler \cite{libre_blog, libre_beck}. This assumption enabled them to remove  much of OpenSSL’s operating system and compiler-specific source code. The LibreSSL team further removed a handful of unnecessary or deprecated ciphers and protocols, many of which dated back to the 1990s, including SSLv2 and Kerberos.

\paragraph{BoringSSL Changes}
BoringSSL has more limited intended use cases than LibreSSL and so was able to discard roughly three times as many LOC as LibreSSL. Since they removed anything not needed for Chromium or Android, BoringSSL discarded a variety of outdated ciphers, protocols, and other algorithms, including Blowfish, Camellia, RC5, MD2, and Kerberos \cite{langley2015}. We also observed that BoringSSL refactored what source code remained more extensively than LibreSSL.

\begin{table}
\centering
\begin{tabular}{@{}lcc@{}}
\toprule
Library   & \begin{tabular}[c]{@{}c@{}}\% of Codebase\\ Removed\end{tabular} & \begin{tabular}[c]{@{}c@{}}\% of Vulnerabilities\\ Removed\end{tabular} \\ \midrule
LibreSSL  & 22\%                                                             & 25\% (15/59)                                                          \\
BoringSSL & 70\%                                                             & 40.6\% (24/59)                                                          \\ \bottomrule
\end{tabular}
\caption{Percentages of source code and CVEs removed in LibreSSL and BoringSSL compared to OpenSSL.}
\label{tab:boring_libre_removed}
\end{table}

\subsubsection{Determining Vulnerabilities Removed}
\label{subsubsection:cves_removed}

The abrupt jettisoning of 22\% and 70\% of OpenSSL’s codebase by LibreSSL and BoringSSL, respectively, raises the question of what impact this had on the security of the two new codebases. Specifically, of the 59 vulnerabilities introduced but not yet discovered in the OpenSSL codebase as of the April 2014 fork, how many still affected LibreSSL and BoringSSL after the steps they took to shrink the codebase? To answer this question, we study vulnerabilities reported in OpenSSL in the wake of Heartbleed and whether they also affected LibreSSL and BoringSSL.

Since we find the official National Vulnerability Database count wholly inaccurate for tracking whether LibreSSL and BoringSSL were affected by CVEs (as described further in \secref{section:cve_reporting}), we create our own database based on OpenSSL’s vulnerability list \cite{openssl_vulnerabilities}. For each CVE affecting OpenSSL post-fork, we manually classify it as having also affected LibreSSL or BoringSSL as of July 11, 2014. To do so, we conducted an extensive manual review of commit descriptions in the source code of both libraries \cite{libre_source, boring_source}, the LibreSSL team mailing list archives \cite{libre_marc}, the LibreSSL change log \cite{libre_changelog}, LibreSSL security advisories \cite{libre_sa}, and the BoringSSL bug tracker \cite{boring_bugs}.

We consider only vulnerabilities introduced prior to the forks (e.g. introduced in OpenSSL version 1.0.1g or earlier) and discovered after July 11, 2014 (e.g. patched in OpenSSL version 1.0.1i or later). We select a July 11 cut-off date since that is the official release date of LibreSSL. BoringSSL was first released in late June, and so by July 11 both libraries had been released. Vulnerabilities discovered and patched between April 7 and July 11 are not included in our dataset to allow the LibreSSL and BoringSSL projects time to make initial modifications to the source code and officially release their versions of the library.

\begin{table}[]
\centering
\begin{tabular}{@{}l|c|c@{}}
\toprule
\multicolumn{1}{c|}{} & \textbf{System} & \textbf{\begin{tabular}[c]{@{}c@{}}Average CCN\end{tabular}} \\ \midrule
\multicolumn{1}{c|}{} & WolfSSL         & 7.8                                                            \\
\multicolumn{1}{c|}{} & OpenSSL         & 6.5                                                            \\
\multicolumn{1}{c|}{} & Libgcrypt       & 6.0                                                            \\
Crypto                & GnuTLS          & 5.9                                                            \\
                      & LibreSSL        & 5.5                                                            \\
                      & Mozilla NSS     & 4.7                                                            \\
                      & BoringSSL       & 3.8                                                            \\
                      & Botan           & 2.2                                                            \\ \midrule
                      & Ubuntu Linux    & 4.3                                                            \\
Non-Crypto            & Wireshark       & 2.9                                                            \\
\multicolumn{1}{c|}{} & Chromium        & 2.2                                                            \\ \bottomrule
\end{tabular}
\caption{Cyclomatic complexities of C/C++ files in cryptographic and non-cryptographic systems, averaged over the previous five major versions for each system.}
\label{tab:ccn_comparison}
\end{table}

In our review of individual project resources, if the team indicated they were not affected by a particular CVE then we use the OpenSSL patch commit location to identify the relevant source code in Libre and Boring and determine why they were unaffected. If the offending OpenSSL source code was not present in the library as of July 11, then we consider the library to have been unaffected by the CVE in the context of our study. Approximately $1/3$ of OpenSSL CVEs were not mentioned in any of the aforementioned LibreSSL or BoringSSL sources, in which case we again revert to independently reviewing the source code covered by the OpenSSL patch and use git diffs to compare files across libraries.

\begin{table*}[]
\centering
\begin{tabular}{@{}l|lllcl@{}}
\toprule
                    & \textbf{System} & \textbf{Versions} & \textbf{LOC Change} & \multicolumn{1}{l}{\textbf{CVEs Introduced}} & \textbf{\begin{tabular}[c]{@{}l@{}}CVEs/\\ KLOC\end{tabular}} \\ \midrule
\textbf{Crypto}     & OpenSSL         & 1.0.0 - 1.0.2     & 58,980              & 70                                           & 1.187                                                         \\
                    & NSS             & 3.12 - 3.21       & 35,554              & 16                                           & 0.450                                                         \\ \midrule
\textbf{Non-Crypto} & Ubuntu Linux    & 10.10 - 15.10     & 3,629,869           & 1,464                                        & 0.403                                                         \\
                    & Wireshark       & 1.4.0 - 2.0.0     & 994,607             & 292                                          & 0.293                                                         \\ \bottomrule
\end{tabular}
\caption{CVEs per thousand lines of code in cryptographic and non-cryptographic systems.}
\label{tab:crypto_noncrypto_ratio}
\end{table*}

\subsubsection{Impact of Code Removal}
Our dataset consists of 59 CVEs introduced in OpenSSL prior to the Heartbleed fork on April 7 and discovered after the releases of both LibreSSL and BoringSSL. Table~\ref{tab:boring_libre_removed} shows that of those 59 CVEs, 44 still affected LibreSSL and just 35 affected BoringSSL. The clear correspondence between the percentage of the OpenSSL codebase removed and the percentage of OpenSSL vulnerabilities removed demonstrates the security implications for reducing codebase size. Figure~\ref{fig:openboringlibre_cves} further breaks down the total based on the year OpenSSL published the CVE. Only the years 2014 through 2016 are included in the figure because no additional CVEs were discovered in 2017 or later that were introduced prior to April 2014 (and therefore qualified for inclusion in our study).

Because these forks inadvertently created a natural experiment in reducing software complexity, we conclude that these vulnerabilities were removed from the respective codebases because the original source code was removed, a stronger conclusion than merely demonstrating a correlation between size and security.

\subsection{Non-Cryptographic Software Complexity}
The findings detailed in \secref{section:complexity} and \secref{section:open_boring_libre} raise the question of whether these findings are unique to cryptographic software. In this section, we study how the relationship between complexity and security in cryptographic systems compares to non-cryptographic systems. We do not study vulnerability type or severity in non-cryptographic systems since there has been an extensive body of work on that subject already \cite{li2017large, ozment2006milk}. Rather, we apply the same methodology used in \secref{section:complexity} to measure complexity in various non-cryptographic systems.

\subsubsection{Cyclomatic Complexity Comparison}
We use McCabe's cyclomatic complexity measure \cite{mccabe1976complexity} as a proxy for estimating the overall complexity of a codebase. We calculate the cyclomatic complexities of all cryptographic libraries studied as well as for three non-cryptographic systems. To vary the systems we selected, we studied an operating system (Ubuntu Linux), a web browser (Chrome/Chromium), and a software application with a security focus (Wireshark).

Table \ref{tab:ccn_comparison} shows the average cyclomatic complexities over the previous five major versions for each of the eight cryptographic libraries studied and the three non-cryptographic systems selected. Six of the eight cryptographic libraries have cyclomatic complexity number (CCNs) that are higher than all three non-cryptographic systems. OpenSSL, in fact, used to have an even higher cyclomatic complexity number of 6.9 in 2014 at the time of Heartbleed as shown in Table~\ref{tab:openssl_versions}. The CCN dropped only after extensive improvements to codebase quality. While this finding does not directly demonstrate subsequent security consequences, it empirically demonstrates the common wisdom that the source code of cryptographic software is significantly more complex than similarly sized non-cryptographic projects.

\subsubsection{CVEs Introduced}
\label{subsubsection:cves_introduced}

To directly study the impact of complexity on vulnerabilities in non-cryptographic systems, we again use a ratio of CVEs reported relative to the number of lines of code introduced. Vulnerability count is too subjective to be used as an absolute metric across systems that vary so widely, and so to more fairly compare security we control for codebase size.

Table \ref{tab:crypto_noncrypto_ratio} reports the number of CVEs per thousand lines of C and C++ source code across two cryptographic systems and two non-cryptographic systems. Of the systems studied, we collect LOC diffs for the major version releases during the six-year period of approximately 2010 through 2015, using 2015 as the cutoff year to allow time for CVEs to be discovered and reported. We further calculate the number of CVEs introduced in those versions, necessarily including only CVEs that accurately listed versions affected. As shown in Table \ref{tab:crypto_noncrypto_ratio}, both OpenSSL and Mozilla NSS have higher rates of CVEs per source code introduced than both non-cryptographic systems, with OpenSSL producing up to three times as many CVEs per line of code.

In selecting systems, we are limited by the quality of the data available. As discussed in \secref{section:systems}, very few systems consistently report the version in which a vulnerability was introduced. Manually tracking down the source code commit for each vulnerability, as we did for Mozilla NSS, is a laborious process and not easily scalable. We exclude Chromium since manual inspection of CVEs reported showed that Chromium, like many systems, reports only the version in which a CVE was patched. We also exclude GnuTLS and other cryptographic libraries for similar reasons.

We further filtered systems by CVSS score on the basis that systems with high-quality vulnerability reporting practices have average and median CVSS scores of around 5.0. For instance, we originally included Android in our study but later excluded it after finding that its vulnerabilities had unusually high CVSS scores, as discussed in \secref{section:cve_reporting}, indicating a strong skew towards reporting only high-severity vulnerabilities.

\subsubsection{Exploitable Lifetime}
The actively exploitable lifetime of a vulnerability helps us further understand security outcomes in software. We collect lifetime data from Ubuntu and Wireshark on CVEs published between 2010 and 2020, inclusive, to compare with the vulnerability lifetimes in cryptographic libraries found in \secref{subsubsection:crypto_lifetimes}.

For Ubuntu, of the 2,187 CVEs studied the average and median lifetimes are 3.89 and 4.03 years, with a standard deviation of 1.44 years. Of the 509 CVEs in Wireshark, the average and median lifetimes are 1.29 and 1.4 years, with a standard deviation of 0.61. We find no comparable difference compared to cryptographic systems and conclude that vulnerability lifetime is mainly a function of a particular project’s version release patterns and patching policies, and tells us little that is generalizable across systems.

\section{Discussion}
\label{section:discussion}
We set out to better understand the connection between software complexity and security and to quantify this connection in cryptographic software. Here, we highlight the most noteworthy results and their implications for software development practices and the ongoing encryption debate.

\medskip
\noindent\textbf{Need for a systems approach to cryptographic software:} Our findings lay bare the discrepancy between the critical role cryptographic libraries hold in securing network traffic and the amount of attention paid to the software quality of the libraries. \secref{subsubsection:code_removal} demonstrated the substantial levels of software bloat in cryptographic software. At the time that Heartbleed was discovered in 2014, OpenSSL had grown to an extent that BoringSSL was able to remove approximately two-thirds of the original codebase while still providing the same core SSL/TLS functionality. Moreover, one of the more interesting trends from Figure~\ref{fig:openboringlibre_size} is that all three of OpenSSL, LibreSSL, and BoringSSL have been gradually increasing in size since the 2014 fork, to the point where LibreSSL is now roughly the same size as OpenSSL was at Heartbleed's discovery. Even projects that set out to be minimalist and security-focused, as both LibreSSL and BoringSSL did, naturally accumulate excess source code and features over time.

This excess is particularly alarming in cryptographic software given that it produces vulnerabilities at higher rates than in non-cryptographic software (as discussed in \secref{subsubsection:cves_introduced}). Since approximately three out of every four vulnerabilities in cryptographic software are caused by common implementation errors, and particularly by memory management issues, overly bloated software threatens significant implications for library security. Furthermore, the most severe vulnerabilities are even \textit{more} likely to have been caused by implementation errors.

All of these findings underscore that the security community needs to devote a level of attention and resources that reflect the importance of cryptographic libraries. Other studies \cite{li2017large, durumeric2014matter} investigating vulnerabilities in cryptographic or open-source software have previously called for our community to improve development and testing processes for security software. Since the scope of this work is to investigate the causes of vulnerabilities in software, not the causes of complexity in software, further research is needed to to understand how to more effectively support OpenSSL and other cryptographic libraries.

\medskip
\noindent\textbf{Don't roll your own crypto:} Our research has demonstrated with empirical analysis of security-critical code bases that that complexity is indeed the enemy of security. While this will not come as a surprise to most in the security community, that fact that we have, for the first time, empirical evidence of just how \textit{much} greater of an impact complexity has on cryptographic software should  garner support for establishing secure practices and standards around cryptographic implementations. The cryptographic libraries we studied produced vulnerabilities at rates of up to one CVE for each thousand lines of code added, a rate roughly three times as much as in non-cryptographic software. In our analysis, we even find one instance where the patch for a vulnerability introduced a new vulnerability \cite{openssl_patch_cve}.

The findings of this work lend empirical support to conventional wisdom preaching the dangers of ``rolling your own crypto''---meaning that software developers should always defer to established libraries and tools instead of re-implementing their own versions. However, cryptography libraries suffer from serious usability issues that make them challenging for non-specialists to navigate. Other studies \cite{nadi2016jumping, acar2017comparing, kruger2017cognicrypt} have found that the software and documentation of these libraries are opaque and that developers overwhelmingly struggle to use them correctly. A usability-centered approach to designing cryptography APIs could make it easier for developers to actually follow this maxim.

\medskip
\noindent\textbf{Lessons of empirical evidence of cryptographic software risk to exceptional access policy:}
Our empirical findings showing high levels of security vulnerabilities in cryptographic codebases have important lessons for the ongoing encryption and surveillance debate. The long-term consequences of general cryptographic source code maintenance provide insights on the ways in which making similar cryptographic changes when creating an exceptional access capability would impact system security.

There are two categories of technical risk that exceptional access has the potential to introduce: (1) protocol design risk and (2) system implementation risk. Much of the historical and contemporary discussion around exceptional access schemes has focused on protocol design: the flaws Blaze uncovered in the NSA's 1993 ``Clipper chip'' proposal \cite{blaze1994protocol} were failures in the underlying transmission protocol. There may have been additional vulnerabilities in the software implementation, but as the cipher was not shared with the public we cannot evaluate those risks. In this work, we evaluated system implementation risk, using vulnerability rates in cryptographic libraries as an approximation of the security consequences of introducing an exceptional access capability.

Recent proposals put forward to implement lawful access have demonstrated an intentional shift away from modifying the underlying cryptography, largely due to anecdotal beliefs in the difficulty of modifying cryptographic code. Levy and Robinson of GCHQ \cite{gchq_principles} stressed in their proposal that a vendor wouldn't have to ``touch the encryption''; Savage \cite{savage} and Ozzie \cite{ozzie_slides} both emphasized that their proposals would not require major new cryptographic primitives; and Wright and Varia \cite{wright2018crypto} highlighted the relative simplicity of their proposed cryptographic constructions. While our findings support the choice to limit the introduction of new primitives, the fact that the majority of vulnerabilities discovered in these codebases were not cryptographic in nature indicate that the overall system complexity (as opposed to the implementation of primitives used) should be considered a more pressing concern; that \emph{protocol and system} complexity in practice overwhelms complexity of cryptographic constructions.

It may be \textit{more} risky to modify cryptographic software, but this should not be taken as evidence that modifying non-cryptographic software is benign. A lawful access scheme requiring non-trivial server and client modifications (as is the case with GCHQ's proposal \cite{gchq_principles}) may yet pose a substantially higher level of risk than a scheme involving minor cryptographic changes. In presenting our findings, we hope to encourage an evidence-based debate about how much security risk a proposal might introduce and whether that level of risk is worth assuming to meet the surveillance needs of law enforcement.

\section{Conclusion}
\label{section:conclusion}
In this paper, we analyzed the impact complexity has on security outcomes in modern cryptographic software, including (1) characteristics of vulnerabilities in cryptographic software and (2) correlations between various complexity metrics and corresponding vulnerability counts. We further presented a comparison of complexity measurements in cryptographic and non-cryptographic systems. We found that only 27.2\% of vulnerabilities introduced in cryptographic software are actually cryptographic, while 37.2\% are memory or resource management issues. Vulnerabilities in cryptographic libraries have an median exploitable lifetime of 4.18 years.

Comparing vulnerability counts in different versions within OpenSSL relative to version size, we find empirical evidence that larger code changes produce more vulnerabilities. In OpenSSL specifically, we find a lower bound of 1 vulnerability introduced for every thousand lines of code. Our case study of the LibreSSL and BoringSSL forks further demonstrates a linear correspondence between source code removal and vulnerability removal. Our findings support the common intuition that it is dangerous to maintain excess amounts of C or C++ source code, and particularly so in cryptographic software.

In our study of non-cryptographic systems, we find that code complexity in cryptographic software is substantially greater than in non-cryptographic software across almost all systems studied, and that non-cryptographic source code generally has a lower density of CVEs introduced compared to cryptographic libraries. Confirming a belief long held within the security community, our findings suggest that cryptographic source code is indeed more brittle and prone to producing security bugs than a comparable amount of source code in a web browser or operating system. The empirical data leads us to conclude that complexity is an even worse enemy of security in cryptographic software than in non-cryptographic software.

\ifanon
\vskip
\else


\fi

\printbibliography

@online{NVD,
  author = {U.S. National Institute of Standards and Technology},
  organization = {National Vulnerability Database},
  addendum = {\url{https://nvd.nist.gov/home.cfm/}}}

@online{CVSS,
  author = {U.S. National Institute of Standards and Technology},
  organization = {CVSS Information},
  addendum = {\url{https://nvd.nist.gov/cvss.cfm/}}}

@online{CVE,
  author = {MITRE Corporation},
  organization = {Common Vulnerabilities and Exposures},
  addendum = {\url{https://cve.mitre.org/}}}

@online{CWE,
  author = {MITRE Corporation},
  organization = {CWE: Common Weakness Enumeration},
  addendum = {\url{https://cwe.mitre.org/}}}

@inproceedings{li2017large,
  title={A Large-Scale Empirical Study of Security Patches},
  author={Li, Frank and Paxson, Vern},
  booktitle={Proceedings of the 2017 ACM SIGSAC Conference on Computer and Communications Security},
  pages={2201--2215},
  year={2017}
}

@inproceedings{shahzad2012large,
  title={A Large Scale Exploratory Analysis of Software Vulnerability Life Cycles},
  author={Shahzad, Muhammad and Shafiq, Muhammad Zubair and Liu, Alex X},
  booktitle={2012 34th International Conference on Software Engineering (ICSE)},
  pages={771--781},
  year={2012},
  organization={IEEE}
}

@inproceedings{ozment2006milk,
  title={Milk or Wine: Does Software Security Improve with Age?},
  author={Ozment, Andy and Schechter, Stuart E},
  booktitle={USENIX Security Symposium},
  volume={6},
  year={2006}
}

@inproceedings{azad2019less,
  title={Less is More: Quantifying the Security Benefits of Debloating Web Applications},
  author={Azad, Babak Amin and Laperdrix, Pierre and Nikiforakis, Nick},
  booktitle={28th USENIX Security Symposium (USENIX Security 19)},
  pages={1697--1714},
  year={2019}
}

@inproceedings{lazar2014does,
  title={Why Does Cryptographic Software Fail? A Case Study and Open Problems},
  author={Lazar, David and Chen, Haogang and Wang, Xi and Zeldovich, Nickolai},
  booktitle={Proceedings of 5th Asia-Pacific Workshop on Systems},
  pages={1--7},
  year={2014}
}

@inproceedings{walden2020impact,
  title={The Impact of a Major Security Event on an Open Source Project: The Case of OpenSSL},
  author={Walden, James},
  booktitle={Proceedings of the 17th International Conference on Mining Software Repositories},
  pages={409--419},
  year={2020}
}

@online{schneier-complexity,
    organization={Schneier on Security},
    title={A Plea for Simplicity: You Can't Secure What You Don't Understand},
    author={Bruce Schneier},
    addendum={\url{www.schneier.com/essays/archives/1999/11/a_plea_for_simplicit.html}},
    year={1999}}

@article{rescorla2005finding,
  title={Is Finding Security Holes a Good Idea?},
  author={Rescorla, Eric},
  journal={IEEE Security \& Privacy},
  volume={3},
  number={1},
  pages={14--19},
  year={2005},
  publisher={IEEE}
}

@inproceedings{zimmermann2010searching,
  title={Searching for a needle in a haystack: Predicting security vulnerabilities for windows vista},
  author={Zimmermann, Thomas and Nagappan, Nachiappan and Williams, Laurie},
  booktitle={2010 Third International Conference on Software Testing, Verification and Validation},
  pages={421--428},
  year={2010},
  organization={IEEE}
}

@article{shin2010evaluating,
  title={Evaluating complexity, code churn, and developer activity metrics as indicators of software vulnerabilities},
  author={Shin, Yonghee and Meneely, Andrew and Williams, Laurie and Osborne, Jason A},
  journal={IEEE transactions on software engineering},
  volume={37},
  number={6},
  pages={772--787},
  year={2010},
  publisher={IEEE}
}

@article{mccabe1976complexity,
  title={A Complexity Measure},
  author={McCabe, Thomas J},
  journal={IEEE Transactions on software Engineering},
  number={4},
  pages={308--320},
  year={1976},
  publisher={IEEE}
}

@online{danial2009cloc,
  title={cloc: Count Lines of Code},
  author={Danial, Al},
  addendum = {\url{https://github.com/AlDanial/cloc}}
}

@online{yin2020lizard,
  organization={Lizard},
  author={Yin, Terry},
  addendum = {\url{https://github.com/terryyin/lizard}}
}

@online{cve2018-16868,
  title={CVE-2018-16868},
  publisher={CVE Details},
  addendum = {\url{https://www.cvedetails.com/cve/CVE-2018-16868/}}
}

@online{cve2020-13777,
  title={CVE-2020-13777},
  publisher={National Vulnerability Database},
  addendum = {\url{https://nvd.nist.gov/vuln/detail/CVE-2020-13777/}}
}

@online{bouncycastle-java,
  title={The Bouncy Castle Crypto Package For Java},
  publisher={Legion of the Bouncy Castle},
  addendum = {\url{https://github.com/bcgit/bc-java}}
}

@online{gaynor2020memory,
  title={What science can tell us about C and C++'s security},
  author={Gaynor, Alex},
  year={2020},
  addendum = {\url{https://alexgaynor.net/2020/may/27/science-on-memory-unsafety-and-security/}}
}

@online{cvedetails,
  organization={CVE Details},
  addendum = {\url{https://www.cvedetails.com/}}
}

@online{opencve,
  organization={OpenCVE},
  addendum = {\url{https://www.opencve.io/}}
}

@online{heartbleed,
   title={The Heartbleed Bug},
   year={2014},
   addendum = {\url{https://heartbleed.com/}}
}

@online{cvssv2v3,
    title={Vulnerability Metrics},
    publisher={National Vulnerability Database},
    addendum = {\url{https://nvd.nist.gov/vuln-metrics/cvss}}
}

@online{libressl,
    organization={LibreSSL},
    addendum = {\url{https://www.libressl.org/}}
}

@online{libre_beck,
    organization={LibreSSl with Bob Beck},
    addendum = {\url{https://www.youtube.com/watch?v=GnBbhXBDmwU}}
}

@online{libre_releases,
    organization={LibreSSL Releases},
    addendum = {\url{https://www.libressl.org/releases.html}}
}

@online{langley2015,
    organization={ImperialViolet},
    title={BoringSSL},
    addendum = {\url{https://www.imperialviolet.org/2015/10/17/boringssl.html}}
}

@online{boringssl,
    organization={BoringSSL},
    addendum = {\url{https://boringssl.googlesource.com/boringssl/}}
}

@online{openssl_vulnerabilities,
    organization={OpenSSL Vulnerabilities},
    addendum = {\url{https://www.openssl.org/news/vulnerabilities.html}}
}

@online{libre_source,
    organization={LibreSSL GitHub},
    addendum = {\url{https://github.com/libressl-portable/openbsd}}
}

@online{boring_source,
    organization={BoringSSL GitHub},
    addendum = {\url{https://github.com/google/boringssl}}
}

@online{libre_marc,
    organization={LibreSSL Mailing list ARChives},
    addendum = {\url{https://marc.info/?l=libressl&r=1&w=2}}
}

@online{libre_sa,
    organization={OpenBSD},
    title={OpenSSL 2015-03-19 Security Advisories: LibreSSL Largely Unaffected},
    addendum = {\url{https://undeadly.org/cgi?action=article&sid=20150319145126}}
}

@online{boring_bugs,
    organization={BoringSSL Bug Tracker},
    addendum = {\url{https://bugs.chromium.org/p/boringssl/issues/list}}
}

@online{libre_changelog,
    organization={LibreSSL ChangeLog},
    addendum = {\url{https://github.com/libressl-portable/portable/blob/master/ChangeLog}}
}

@online{libre_blog,
    author={Ted Unangst},
    title={LibreSSL: More Than 30 Days Later},
    addendum = {\url{https://www.openbsd.org/papers/eurobsdcon2014-libressl.html}}
}

@online{censys,
    organization={Censys},
    addendum = {\url{https://censys.io/}}
}

@online{openssl_releases,
    organization={OpenSSL Releases},
    addendum = {\url{https://github.com/openssl/openssl/releases}}
}

@online{openssl,
    organization={OpenSSL},
    addendum = {\url{https://www.openssl.org/}}
}

@online{gnutls,
    organization={GnuTLS},
    addendum = {\url{https://www.gnutls.org/}}
}

@online{nss,
    organization={Mozilla Network Security Services},
    addendum = {\url{https://developer.mozilla.org/en-US/docs/Mozilla/Projects/NSS}}
}

@online{botan,
    organization={Botan},
    addendum = {\url{https://botan.randombit.net/}}
}

@online{libgcrypt,
    organization={Libgcrypt},
    addendum = {\url{https://gnupg.org/software/libgcrypt/index.html}}
}

@online{wolfssl,
    organization={WolfSSL},
    addendum = {\url{https://www.wolfssl.com/}}
}

@online{ubuntu,
    organization={Ubuntu Linux},
    addendum = {\url{https://ubuntu.com/}}
}

@online{chrome,
    organization={Chromium},
    addendum = {\url{https://www.chromium.org/Home}}
}

@online{wireshark,
    organization={Wireshark},
    addendum = {\url{https://www.wireshark.org/}}
}

@inproceedings{durumeric2014matter,
  title={The Matter of Heartbleed},
  author={Durumeric, Zakir and Li, Frank and Kasten, James and Amann, Johanna and Beekman, Jethro and Payer, Mathias and Weaver, Nicolas and Adrian, David and Paxson, Vern and Bailey, Michael and others},
  booktitle={Proceedings of the 2014 conference on internet measurement conference},
  pages={475--488},
  year={2014}
}

@online{cryptlib,
    organization={CryptLib},
    addendum = {\url{https://www.cryptlib.com/}}
}

@online{cvss_scoring,
    organization={National Vulnerability Database},
    title={CVE FAQs},
    addendum = {\url{https://nvd.nist.gov/general/FAQ-Sections/CVE-FAQs\#\#faqLink10}}
}

@online{openssl_spike,
    organization={OpenSSL GitHub},
    title={aes ctr\_drbg: add cavs tests},
    addendum = {\url{https://github.com/openssl/openssl/commit/e613b1eff40f21cd99240f9884cd3396b0ab50f1}}
}

@online{openssl_patch_cve,
    organization={OpenSSL Security Advisory [26 Sep 2016]},
    addendum = {\url{https://www.openssl.org/news/secadv/20160926.txt}}
}

@online{gchq_principles,
  title={{Principles for a More Informed Exceptional Access Debate}},
  author={Levy, Ian and Crispin Robinson},
  howpublished={{Lawfare Blog}},
  year={2018},
}

@conference{savage,
   author={Stefan Savage},
   year={2018},
   title={Lawful device access without mass surveillance risk: A technical design discussion},
   publisher={Proceedings of the 2018 ACM SIGSAC Conference on Computer and Communications Security},
}

@misc{ozzie_slides,
  title={{CLEAR}},
  author={Ray Ozzie},
  howpublished={\url{https://github.com/rayozzie/clear/blob/master/clear-rozzie.pdf}},
  year={2017},
}

@inproceedings{wright2018crypto,
  title={Crypto Crumple Zones: Enabling Limited Access Without Mass Surveillance},
  author={Wright, Charles and Varia, Mayank},
  booktitle={2018 IEEE European Symposium on Security and Privacy (EuroS\&P)},
  pages={288--306},
  year={2018},
  organization={IEEE}
}

@inproceedings{blaze1994protocol,
  title={Protocol Failure in the Escrowed Encryption Standard},
  author={Blaze, Matt},
  booktitle={Proceedings of the 2nd ACM Conference on Computer and Communications Security},
  pages={59--67},
  year={1994}
}

@article{keysunderdoormats,
   author={Abelson, Harold and Ross Anderson and Steven M. Bellovin and Josh Benaloh and Matt Blaze and Whitfield Diffie and John Gilmore and Matthew Green and Susan Landau and Peter G. Neumann and Ronald L. Rivest and Jeffrey I. Schiller and Bruce Schneier and Michael Specter and Daniel Weitzner},
   title={Keys Under Doormats: Mandating Insecurity by Requiring Government Access to All Data and Communications},
   journal={Journal of Cybersecurity},
   volume={1.1},
   year=2015,
   pages={69-79}
}

@article{chang2013your,
  title={Is Your Data on the Healthcare. gov Website Secure?, Written Testimony before the Committee on Science},
  author={Chang, Frederik},
  journal={Space and Technology, US House of Representatives, November},
  year={2013}
}

@inproceedings{nadi2016jumping,
  title={Jumping Through Hoops: Why Do Java Developers Struggle With Cryptography APIs?},
  author={Nadi, Sarah and Kr{\"u}ger, Stefan and Mezini, Mira and Bodden, Eric},
  booktitle={Proceedings of the 38th International Conference on Software Engineering},
  pages={935--946},
  year={2016}
}

@inproceedings{acar2017comparing,
  title={Comparing the Usability of Cryptographic APIs},
  author={Acar, Yasemin and Backes, Michael and Fahl, Sascha and Garfinkel, Simson and Kim, Doowon and Mazurek, Michelle L and Stransky, Christian},
  booktitle={2017 IEEE Symposium on Security and Privacy (SP)},
  pages={154--171},
  year={2017},
  organization={IEEE}
}

@inproceedings{kruger2017cognicrypt,
  title={Cognicrypt: Supporting Developers in Using Cryptography},
  author={Kr{\"u}ger, Stefan and Nadi, Sarah and Reif, Michael and Ali, Karim and Mezini, Mira and Bodden, Eric and G{\"o}pfert, Florian and G{\"u}nther, Felix and Weinert, Christian and Demmler, Daniel and others},
  booktitle={2017 32nd IEEE/ACM International Conference on Automated Software Engineering (ASE)},
  pages={931--936},
  year={2017},
  organization={IEEE}
}

\end{document}